# Enhancement of light emission in Bragg monolayer-thick quantum well structures

Galia Pozina[1*], Konstantin A. Ivanov[2], Konstantin M. Morozov[3], Elizaveta I. Girshova[3], Anton Yu. Egorov[2,4], Stewart J. Clark[5], Mikhail A. Kaliteevski[2,3,4]

[1] *Department of Physics, Chemistry and Biology (IFM), Linköping University, S-581 83 Linköping, Sweden*

[2] *ITMO University, Kronverkskiy pr. 49, 197101 St. Petersburg, Russian Federation*

[3] *St-Petersburg Academic University, Khlopina 8/3, 194021 St. Petersburg, Russian Federation*

[4] *Ioffe Institute, Politekhnicheskaya 26, 194021 St. Petersburg, Russian Federation*

[5] *Department of Physics, Durham University, South Road, Durham, UK DH1 3LE*

\* galia.pozina@liu.se

Control over spontaneous emission rate is important for improving efficiency in different semiconductor applications including lasers, LEDs and photovoltaics. Usually, an emitter should be placed inside the cavity to increase spontaneous emission rate, although it is technologically challenging. Here we experimentally demonstrate a phenomenon of super-radiance observed in a cavity-less periodic Bragg structure based on InAs monolayer-thick multiple quantum wells (MQW). The collective super-radiant mode shows enhanced emission rate for specific angles and frequencies. This behaviour correlates with the calculations demonstrating individual spots of enhanced Purcell coefficient near the Bragg condition curve. This study provides a perspective for realization of surface emitting cavity-less lasers with distributed feedback.

**Introduction**

Mastering the modification of the radiation emission rates paves the way to a realization of new functionality in optoelectronic devices. The principal way to enhance these rates is to place an emitter in a strongly structured optical medium, which leads to the alteration of the photon density of states (DOS). A direct consequence of this, as follows from the Fermi's golden rule, is so-called Purcell effect[1-3]. Purcell suggested that the control of the spontaneous emission rate of a quasi-monochromatic dipole can be done in a cavity[2]. Then the spontaneous emission rate at the resonance frequency with the cavity mode will be enhanced, while emission at other non-resonance frequencies will be suppressed. However, the technological challenge is to integrate such resonators in devices demonstrating considerable output power since experimental realization requires cavities with high Q-factors and small mode volumes.

On the other hand, the spectacular phenomena of the super-radiance (SR) requires no cavity but only an ensemble of two-level atoms occurring in symmetric Dicke states to obtain an enhanced emission rate[4,5]. Remarkably, atoms in these states show a rate of spontaneous emission proportional to the square of the number of emitters N. Such super-radiant behaviour of the emission is a result of the strong quantum correlations among the emitters occurring in symmetric Dicke states[6]. However, it is difficult to prepare a highly entangled states for a large number of particles. From this point of view, a revolutionary technological breakthrough can be achieved for multiple quantum well (MQW) structures based on mature III-V semiconductors. Theoretically it was shown by Ivchenko et al.[7] that such interaction can occur for the Bragg arrangement of MQW structures, when the period of the structure $d$ is coupled to the exciton emission frequency $\omega$ by the Bragg condition,

$$n\omega d \cos\varphi = \pi c \qquad (1)$$

where $\varphi$ is the angle of propagation in the structure, and $n$ is the refractive index of the structure. The angle of emission into the vacuum $\theta$ is described by

$$\omega d\sqrt{n^2 - \sin^2\theta} = \pi c \qquad (2)$$

The first experiments with Bragg MQWs have demonstrated the resonant reflection coefficient of ~7 %[8], while for the same but single semiconductor QW the magnitude of the exciton resonance in reflection spectra was less than 1 %[9]. Advances in epitaxial technology have allowed fabrication of highly coherent III-V semiconductor-based Bragg MQW structures demonstrating excitonic reflection above 50 %[10-12]. Recent studies of luminescence properties in Bragg MQW structures grown by molecular beam epitaxy (MBE), formed by InAs monolayers confined in GaAs matrix, has demonstrated a clear appearance of an additional blue-shifted collective emission mode revealing a super-radiant behaviour. It was preliminary assigned to the super-radiant emission of coherently coupled individual quantum well excitons[12,13].

However, to make a decisive conclusion about the nature of additional mode observed[12,13], we prove here that the new mode with super-radiant behaviour is in fact the result of coherent interference between the emissions from individual quantum wells and is not associated with the recombination of hot nonequilibrium carriers. Furthermore, we give insight into fundamental problems regarding the modification of the spontaneous emission rate in InAs single monolayer MQWs with GaAs barriers using a detailed analysis of a modification of the spontaneous emission rate in such systems. Our experimental and theoretical results lead not only to an understanding of the optical properties in Bragg MQW structures but also pave the way to novel potential applications, ranging from optical logic devices and optical switches to the realization of highly interesting surface emitting cavity-less lasers with distributed feedback. These also combine advantages such as low divergence typical for VCSEL[14] with high monochromatic properties typical for distributed feedback lasers[15,16].

**Mixing of exciton states in the triple monolayer QW structure**

The structure for the experimental studies is illustrated in Fig. 1. Also, the geometry of the excitation and detection of the emission is indicated. The sample grown by MBE consists of 60 triple InAs monolayer quantum wells, where InAs monolayers are separated by 10 nm of GaAs, while the triple quantum wells are separated from each other by 102 nm of GaAs. The eigenenergies and wavefunction of electron localized at the InAs monolayer have been calculated using density functional theory (DFT)[17] implemented in the CASTEP computer code[18]. For more details about the experiment conditions, the structure preparation and the calculations see Methods.

First, we consider the carrier localization in the InAs monolayer embedded in the GaAs matrix. In such QW system, the localization energy for electrons and holes is 16 meV and 19 meV, respectively. Thus, for the exciton localized at the single InAs monolayer QW the energy in the ground state is 1.480 eV at temperature of 4 K. The charge density is shown in Fig. 2a and b (top) with corresponding profiles (bottom) of the probability density for the holes and electrons localized at one monolayer of InAs confined by GaAs, respectively. It can be seen that for electrons, the localization length ($a$) along z-direction is about 10 nm, while for the hole it is 5 nm. Thus, due to a relatively long localization length of electrons we have an overlapping of the wave function of electrons in the triple QWs, which will result in a triplet structure of the exciton mode with eigen frequencies near 1.471 eV, 1.482 eV and 1.492 eV as described in details in [12]. Such mixing of exciton states can be seen as a peculiar triplet structure in the both experimental (blue line) and modelled (red line) reflection spectrum of the Bragg monolayer QW structure shown in Fig. 3.

Peculiarities in the reflection spectra have been studied with respect to the dependence on light polarization for different incidence angles. The reflection coefficient calculated as a function of incidence angle and photon energy for the Bragg MQW structure is shown in Fig. 3a and d for TE and TM-polarization, respectively. Bragg conditions are indicated by a solid magenta line. The dashed lines in Fig. 3a and d correspond to the incidence angles of 45° and 65°, close to Brewster angle of ~72° for GaAs. A comparison of the calculated reflection curves with experimental reflection spectra for both TE and TM polarizations is done for 45° and 65° as shown in Fig. 3b, c and e, f, respectively. The features in the experimental and calculated spectra correlate very well and the energy position of the peaks follow Bragg conditions.

**Additional super-radiant emission mode**

Results of the time-resolved photoluminescence (TRPL) measured from the top of the sample has shown that under increasing of pumping an additional mode appears with a peak energy depending on the angle. The dependence of the intensity of this mode is proportional to nearly the squared pumping power[12,13], therefore, this mode is likely originated because of SR[4].

Also, the time delay between the central mode (X1) and the SR mode is very clear with the increase of excitation power, while the recombination time became much faster. At the same time, measurements performed on the edge of the structure do not show the appearance of the additional mode. We demonstrate this in Fig. 4 for the TRPL measured at different power for an angle of 40° (Fig. 4a, b, c) and compare to the same measurements performed at the edge of the structure (Fig. 4d, e, f).

We point out the following features in the case of emission from the edge: there is only a strong emission line corresponding to ground state exciton (X1) with the peak energy of ~1.47 eV; there is no change in the delay between the exciting pulse and luminescence for different excitation powers; and there is no change in the luminescence decay time for different excitation powers. All these results indicate that there is no super-radiance or non-linear effects in the emission taken from the edge. In contrast, the emission from the top of the sample significantly changes with increasing emission power (Fig. 4a, b, c).

To understand the properties of the SR mode, it is convenient to perform a quantitative comparison of the emission pattern from the top and from the edge of the sample. For this purpose, we have measured the TRPL for different angles in the interval between 15° and 85° with the step of 2° (see examples of experimental data taken for different angles in Supplementary Fig. S3). The peak position of the SR mode varies in time due to a change of the refractive index of the material caused by the recombination of non-equilibrium electrons and holes generated by the excitation pulse.

In order to make a quantitative analysis of influence of the Bragg arrangements of quantum wells on probability of spontaneous emission, we have applied the following procedure:

i) we integrated the signal of TRPL *F* over time within a specific time interval (as shown in Fig. 4c and f by horizontal lines) to obtain time-integrated intensities for the emission taken from the sample top:

$$I_s(\omega, \theta) = \int F_s(\omega, t, \theta) dt \qquad (3)$$

and from the edge of the sample:

$$I_e(\omega) = \int F_e(\omega, t) dt. \qquad (4)$$

ii) we then obtained normalized intensities

$$\widetilde{I_s}(\omega, \theta) = I_s(\omega, \theta) / \int I_s(\omega, \theta) d\omega \qquad (5)$$

and

$$\widetilde{I_e}(\omega) = I_e(\omega) / \int I_e(\omega) d\omega; \qquad (6)$$

iii) and finally, we got the ratio:

$$P(\omega,\theta) = \tilde{I}_s(\omega,\theta)/\tilde{I}_e(\omega) \ . \qquad (7)$$

The procedure of integration over a finite time interval is required to smooth the noise in the TRPL signal. Since the emission from the edge of the sample is not affected by the Bragg arrangement of monolayer quantum wells, $P(\omega,\theta)$ is the experimental counterpart of the ratio of spontaneous emission rates for a particular direction for the emitter placed into the structure and for the emitter in free space (i.e. the modal Purcell factor)[19].

Fig. 5 shows the pattern of $P(\omega,\theta)$ for different time integration intervals, where time starts from the excitation laser pulse. The Bragg condition (solid magenta line) is shown on each pattern for comparison. Fig. 5a shows $P(\omega,\theta)$ for the time integration interval from 0 to 700 ps, which corresponds to the time of existence of the SR mode in the spectra. $P(\omega,\theta)$ is enhanced for the frequency of light corresponding to the ground state of exciton and also along the Bragg condition line. Since the peak position of the SR mode is shifted with time, it is convenient to reduce the time integration interval. Thus, Fig. 5b corresponds to the time integration interval between 100 ps and 200 ps, when only the SR mode exists in the spectra (see Fig. 4c). It is clearly seen, especially in Fig. 5b, that the enhancement of the emission occurs only for some spots along the Bragg condition line. For the time integration interval between 200 ps and 300 ps (Fig. 5c) and between 300 ps and 400 ps (Fig. 5d) the spotted pattern of $P(\omega,\theta)$ along the Bragg condition is conserved, however the spots become less pronounced and instead the feature corresponding to the X1 line appears. For the integration intervals between 500 ps and 600 ps and also between 600 ps and 700 ps shown in Fig. 5e and f, respectively, the SR mode decays and the enhancement of the emission along Bragg condition vanishes.

**Modal Purcell factor calculations**

The spotted structure in the emission enhancement pattern is the most intriguing and counter–intuitive finding of this work: an enhancement of spontaneous emission due to super-radiance is expected to occur for all directions and frequencies of the emission coupled to the Bragg condition equation (1). Fig. 5b and c shows that the enhancement of the emission occurs only for discrete periodically placed spots near the Bragg condition.

Insight on such unusual behaviour can be obtained by analysis of the modal Purcell factor[19]. Fig. 6 shows examples of the modal Purcell factor calculated for different positions of the emitter placed inside the structure. It can be seen that there is an area of enhanced modal Purcell factor near the bottom of the Bragg condition curve. Also, there are individual spots of enhanced Purcell factor positioned near the Bragg condition curve at the frequencies corresponding to the triplet exciton states. Such behaviour correlates with the experimental pattern of the emission enhancement. Thus, the observed spotted TPRL pattern results from the combination of several effects: (i) the Purcell enhancement of spontaneous emission rate; (ii) super-radiance; (iii) spatial diffusion, and (iv) the relaxation of non-equilibrium electrons and holes.

**Conclusions**

A Bragg quantum well structure based on MBE-grown InAs monolayers with GaAs barriers has been fabricated and studied. Eigenenergies and wavefunctions of electron and holes localized on InAs monolayers have been calculated by a density functional approach. Time-resolved photoluminescence of the Bragg quantum well structure has been measured from the edge of the sample (where light emission is not affected by Bragg arrangement of quantum wells) and from the surface of the sample for different emission angles. We have shown that emission from the edge of the sample is represented by one line corresponding to the ground state exciton, while for the emission from the top of the sample an additional line appears and the angle of emission and the frequency of light for this line corresponds to Bragg conditions. The ratio of time-integrated emission intensities from the surface of the sample for different angles and from the edge of the samples as a function of frequency of light has been obtained, showing the pattern of this ratio has a spot-like structure, where spots are located near the Bragg condition. Such behaviour correlates with the calculated dependence of the modal Purcell factor on the frequency and angle of emission, which also demonstrates individual spots of enhanced Purcell coefficient near the Bragg condition curve.

**Methods**

*Sample growth*

The Bragg structure contained 60 triple QWs based on monolayer-thick InAs separated by GaAs layers. The thickness of layers was chosen to satisfy the Bragg conditions for the emission wavelength corresponding to the exciton emission at 5 K in the single monolayer InAs QW with GaAs barriers. This matches to the peak photon energy of 1.48 eV, while the optical thickness of the period in the Bragg structure is 418 nm. Each triple QWs was separated by the 102 nm-thick undoped GaAs layer, while the spatial separation between individual InAs monolayers in the triple QW was 10 nm. The sample was fabricated under rotation using the molecular beam epitaxy growth chamber Riber 49. (100)-GaAs was used as the substrate. High energy electron diffraction was used to control in-situ the growth rate. The thickness of the GaAs cap layer was about 100 nm to provide a matching of the antinodes of the electric field of the collective mode. The part of the GaAs cap layer near the sample surface was doped in order to avoid the GaAs excitonic reflection feature from the surface of the sample.

*Characterization*

Time-integrated and time-resolved photoluminescence was studied under excitation by a Ti: sapphire femtosecond pulsed laser with a frequency of 75 MHz. The excitation wavelength was 800 nm having a maximum average power density of 100 W/cm$^2$. The excitation pulse duration was ~150 fs.

A Hamamatsu syncroscan streak camera with a temporal resolution of ~20 ps was employed to acquire the TRPL signal. The sample was cooled down to 5 K using a liquid-He cryostat. The angle of emission was tuned by rotating the sample holder.

*Modelling*

The calculated dependence of the reflection coefficient on the angle of emission and frequency of light was obtained using transfer matrix method, detailed description of the method is provided in [12] (see also Supplementary equations (S1-S9)). Dependence of probability density of spontaneous emission on the angle of emission and frequency of light was obtained by S-quantization formalism described in detail in [19] and in Supplementary equations (S10-S41).

*Electronic structure calculations*

The electronic structure calculations were performed using the CASTEP code[17] which uses the plane wave density functional formalism within the generalised gradient approximation.

**Data availability**. All data generated and/or analysed during this study are available from the corresponding author on reasonable request.

**Acknowledgements**
The work has been supported by Russian Science Foundation Grant 16-12-10503. SJC acknowledges Durham's Hamilton and UK's Archer supercomputer facilities for the DFT calculations.

**Author contributions**
GP, MAK designed the research idea. GP performed optical measurements. KAI, KMM, MAK, AYE contributed to the design of the sample. EIG, KAI, SJC and KMM contributed to numerical modelling of the results. All authors contributed to analysis of the results. GP, KAI, MAK wrote the manuscript. All authors have given approval to the final version of the manuscript.

**Additional Information**
Competing interests: The authors declare no competing interests.

**References**

1. Kleppner, D. Inhibited spontaneous emission. *Phys. Rev. Lett.* **47**, 233-236 (1981).
2. Purcell, E. M., Torrey, H. C. & Pound, R. V. Resonance Absorption by Nuclear Magnetic Moments in a Solid. *Physical Review* **69**, 37-38 (1946).
3. Gérard, J. *et al.* Enhanced spontaneous emission by quantum boxes in a monolithic optical microcavity. *Phys. Rev. Let.* **81**, 1110-1113 (1998).
4. Dicke, R. H. Coherence in spontaneous radiation processes. *Phys. Rev.* **93**, 99-110 (1954).
5. Rehler, N. E. & Eberly, J. H. Superradiance. *Phys. Rev. A* **3**, 1735-1751 (1971).
6. Gross, M. & Haroche, S. Superradiance: An essay on the theory of collective spontaneous emission. *Phys. Reports* **93**, 301-396 (1982).
7. Ivchenko, E., Nesvizhskii, A. & Jorda, S. Bragg reflection of light from quantum-well structures. *Phys. Sol. State* **36**, 1156-1161 (1994).
8. Kochereshko, V. P. *et al.* Giant exciton resonance reflectance in Bragg MQW structures. *Superlat. Microstr.* **15**, 471-473 (1994).
9. Ivchenko, E. L. *et al.* Reflection in the exciton region of the spectrum of a structure with a single quantum well. Oblique and normal incidence of light. *Sov. Phys. Semicond.* **22**, 495-498 (1988).
10. Hübner, M. *et al.* Optical Lattices Achieved by Excitons in Periodic Quantum Well Structures. *Phys. Rev. Let.* **83**, 2841-2844 (1999).
11. Chaldyshev, V. V. *et al.* Optical lattices of InGaN quantum well excitons. *Appl. Phys. Let.* **99**, 251103 (2011).
12. Pozina, G. *et al.* Super-radiant mode in InAs—monolayer–based Bragg structures. *Sci. Rep.* **5**, 14911 (2015).
13. Pozina, G. *et al.* Nonlinear behavior of the emission in the periodic structure of InAs monolayers embedded in a GaAs matrix. *Phys. Stat. Sol. b* **254**, 1600402 (2017).
14. Khurgin, J. B. & Sun, G. Comparative analysis of spasers, vertical-cavity surface-emitting lasers and surface-plasmon-emitting diodes. *Nature Phot.* **8**, 468-473 (2014).
15. Turitsyn, S. K. *et al.* Random distributed feedback fibre laser. *Nature Phot.* **4**, 231-235 (2010).
16. Wang, Z. *et al.* Room-temperature InP distributed feedback laser array directly grown on silicon. *Nature Phot.* **9**, 837-842 (2015).
17. Clark, S. J. *et al.* First principles methods using CASTEP. *Z. Kristal.* **220**, 567-570 (2005).
18. Milman, V. *et al.* Electron and vibrational spectroscopies using DFT, plane waves and pseudopotentials: CASTEP implementation. *J. Molec. Struc.: THEOCHEM* **954**, 22-35 (2010).
19. Gubaydullin, A. R. *et al.* Enhancement of spontaneous emission in Tamm plasmon structures. *Sci. Rep.* **7**, 1914 (2017).

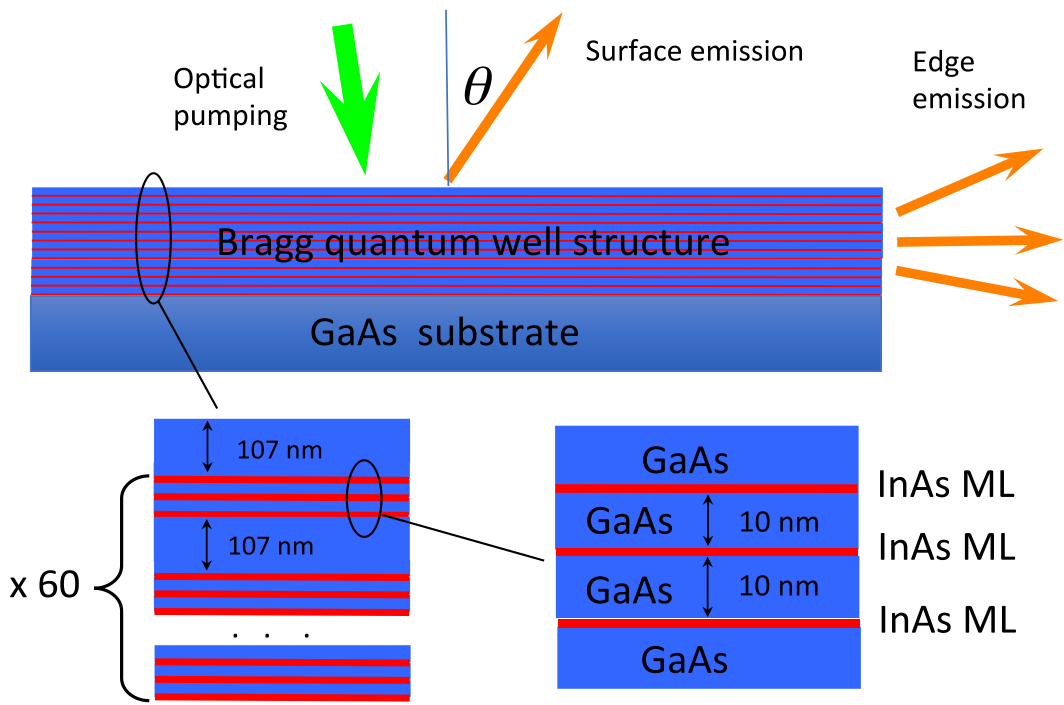

**Figure 1**. Schematic draws of the experimental structure and the measurements of photoluminescence.

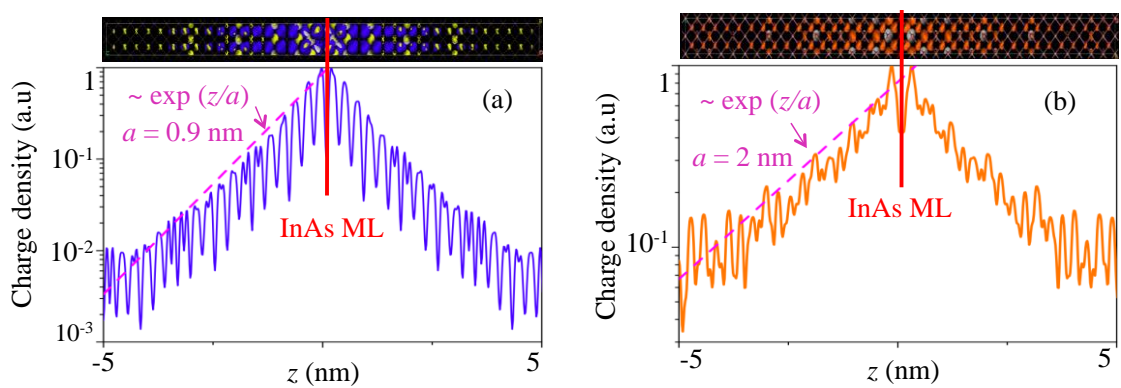

**Figure 2.** Calculated charge density. Averaged in the plane of InAs monolayer charge density of holes (a) and electrons (b) localized at InAs monolayer confined by GaAs. The coordinate z corresponds to the structure growth axis. On the top - images of charge density are shown.

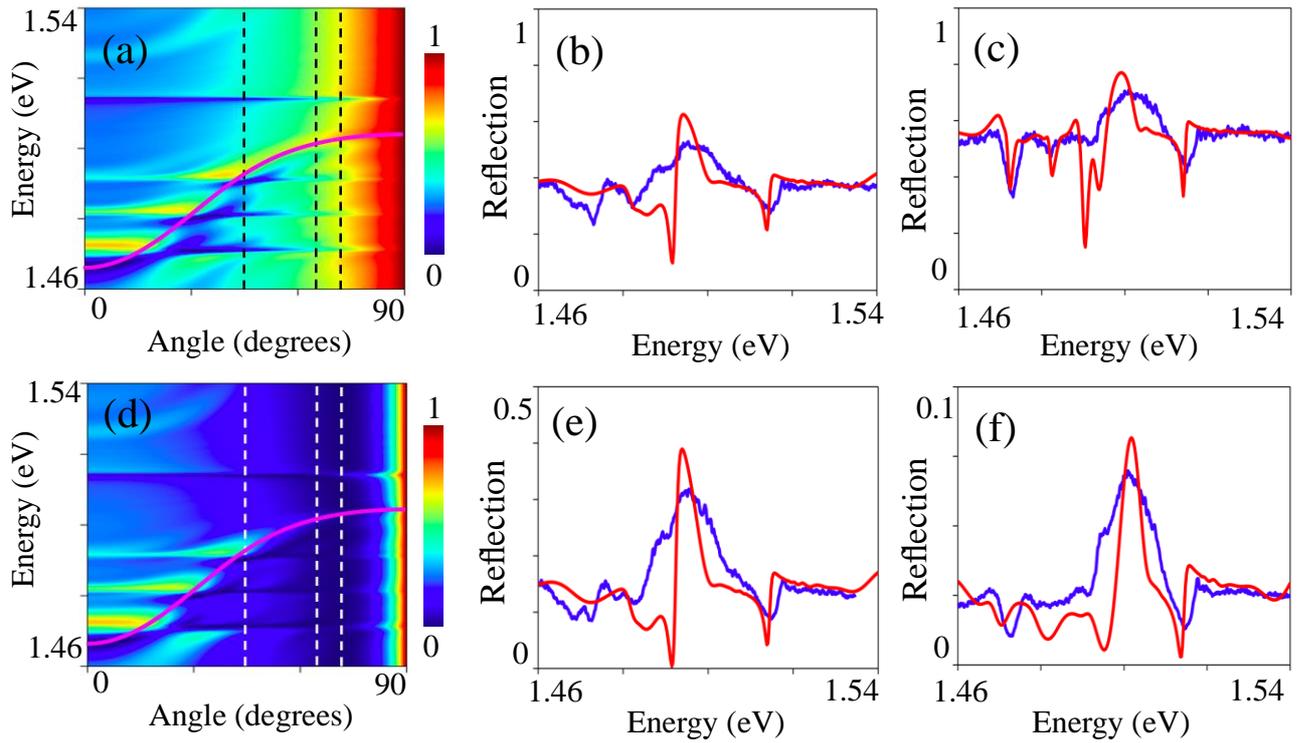

**Figure 3**. Reflection for the studied sample. Calculated reflection coefficient as a function of incidence angle and photon energy for (a) TE – polarization and (d) for TM-polarization. Bragg condition is shown by the magenta solid line. Vertical dashed lines in (a, d) indicate the angle of incidence of 45°, 65° and 72° (Brewster angle). Experimentally measured (blue lines) and calculated (red lines) spectra in TE -polarization for an incidence angle of 45° (b) and 65° (c). Experimental (blue lines) and calculated (red lines) reflection spectra in TM-polarization for the angles of 45° (e) and 65° (f). These calculations are based on the transfer matrix method, see Supplementary equations (S1-S9) with parameters given in Supplementary Table S1.

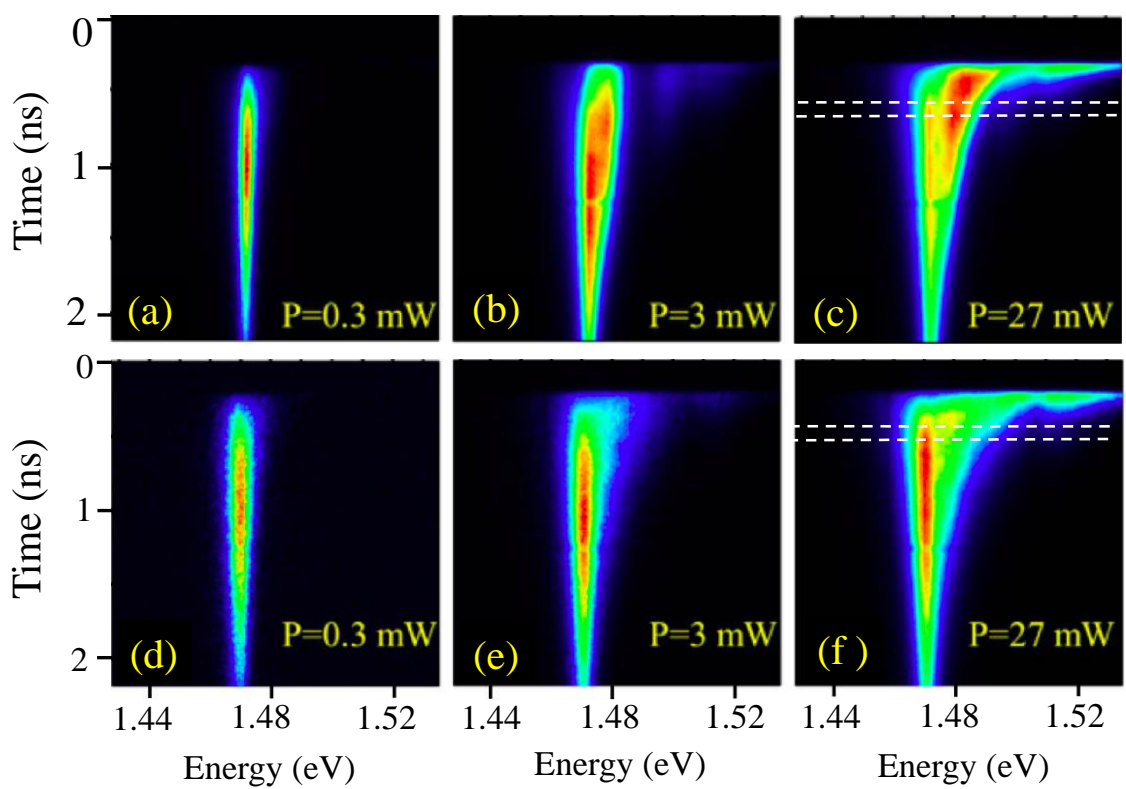

**Figure 4**. Transient photoluminescence pattern. (a,b,c) -measurements from the surface of the sample at the emission angle of 40°, (d,e,f) measurements from the edge of the sample. Pumping power P used 0.3 mW (a, d); 3 mW (b, e); and 27 mW (c, f). Horizontal dashed lines indicate an example of interval of integration in time (200 ps -300 ps). Examples of PL spectra taken at different delay time is shown in Supplementary Fig. S1.

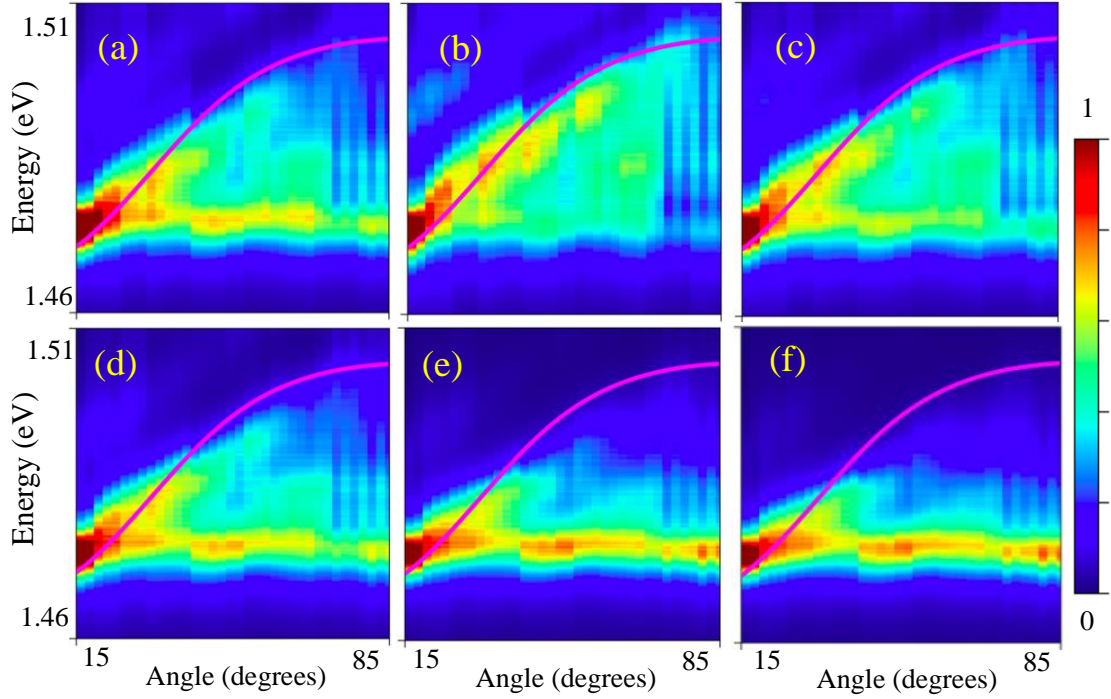

**Figure 5**. Dependence of the ratio of time –integrated emission spectra taken from surface and from the edge of the sample (P(ω, θ)) obtained by equation (3) for different time integration intervals: (a) 0 ps-700 ps; (b) 100 ps -200 ps; (c) 200 ps-300 ps; (d) 300 ps-400 ps; (e)500 ps -600 ps; and (f) 600 ps -700 ps. Time interval is calculated from the start of excitation pulse. Time-integrated PL spectra measured for different angles are shown in Supplementary Fig. S2.

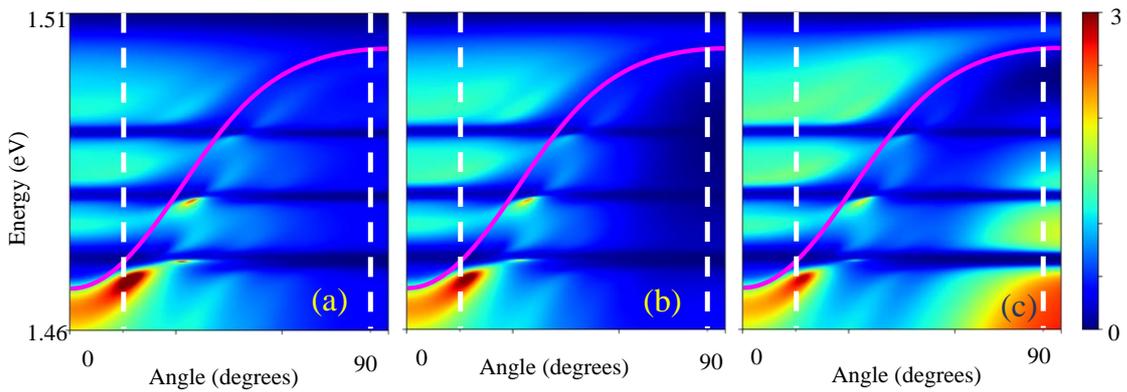

**Figure 6**. Dependence of calculated modal Purcell factor on the angle and photon energy for TE-polarization when emitter is placed in (a) central monolayer in the 1$^{st}$ period; (b) central monolayer in the 5th period; (c) central monolayer in the 10th period. Vertical dashed lines indicate the interval of angles from 15° to 85°, for which the experimental dependences of the emission in Fig. 5 are shown. Calculations are based on S-quantization formalism (Methods), (for details see Supplementary equations (S10-S41)).